# *On the possible physical mechanism of Chernobyl catastrophe and the unsoundness of official conclusion*

**A.A. Rukhadze,* L.I. Urutskojev,** D.V. Filippov****

** General Physics Institute, Russian Academy of Science*
*** RECOM, Russian Research Center «Kurchatov Institute»*
*e-mail: recom@hotmail.ru, shevchenko_e@mail.ru*

**Abstract**

The official conclusion about the origin and mechanism of the Chernobyl catastrophe is shown to essentially contradict experimental facts available from the accident. In the frame of existing physical models of nuclear fission reactor, it is shown analytically that under conditions of the accident the period of runaway of reactor at the fourth power generating unit of the Chernobyl Nuclear Power Plant (CNPP) should be either 10 times slower or 100 times faster than that observed. A self-consistent hypothesis is suggested for the probable birth of magnetic charges, during the turbine generator test under it's own momentum test, at the fourth power generating unit of CNPP, and for the impact of these charges on the reactivity coefficient.

## 1. INTRODUCTION

The reason that has incited the author to the work on this publication is strong conviction that without gaining full understanding of the mechanism of one tragedy, we shall sooner or later witness another one. The official conclusion seems to be unsatisfactory: first, many questions, as shown below, have not been answered and, second, the official conclusion was based on a numerical modeling whose results do not agree with our analytic estimations.

This paper is an attempt to put forward a rather "exotic" hypothesis as the main physical mechanism of the accident. According to this hypothesis, magnetic monopoles are generated during the turbine generator test under it's own momentum and enter the nuclear reactor together with the steam. Despite the apparent unreason of such a hypothesis, it provides a simple and logical interpretation for a variety of experimental facts that could not be explained previously.

The possibility of existence of magnetic charges in nature has been widely discussed in physics since the publication of Paul Dirac's paper [1]. Numerous failures in experimental detection of magnetic charges considerably abated the researchers' enthusiasm. However, the construction of the Grand Unification Theory (GUT) [2, 3] has revived the ideas of so-called "heavy" monopoles or GUT-monopoles. According to the modern views, magnetic monopoles could be formed at early stages of the Universe evolution and only so-called "relic" monopoles could persist till now. A set of very interesting effects have been predicted theoretically for the GUT-monopoles [4], in particular, nucleon disruption [5], i.e. a process accompanied by violation of the baryonic charge conservation law. About 20 years ago, studies by George Lochak [6,7] were published in which the magnetic monopole was derived from solving the Dirac equation. Lochak [6] found that the Dirac equation in Weyl's representation allows a pseudo-vector class of solutions; thus permitted him to derive the second pair of Maxwell's equations. By expressing the magnetic charge pseudo-scalarity in terms of the $\boldsymbol{\gamma_5 g}$ matrix [7], instead of the $\mathbf{g}$ constant for magnetic charge, one can represent the electromagnetism as a "mountain having two slopes – a vector (electromagnetic) and a pseudo-vector (magnetoelectric) ones". The magnetic charge introduced by Lochak obeys the Schwinger quantization condition [8, 6]; it is a lepton, i.e. a participant of electroweak interactions, and can be interpreted as a magnetically excited state of neutrino. This monopole is massless (or almost massless), i.e., it is very light (from the energy standpoint) and can be produced during electromagnetic phenomena [9].

## 2. THE QUESTIONS NOT ANSWERED

In the authors' opinion, a number of experimental facts that stimulated musing on the root of the Chernobyl accident have not been explained convincingly. These include:

- The mechanism of the reactor acceleration: how did it come that the reactor with a high rate of fuel consumption (up to 20 MW day/kg) and spoiled with xenon was accelerated, in 10 seconds, from 200 MW power level (i.e. 6% of nominal power) to the level exceeding the nominal by a factor of several tens;
- The integrity of structures in the reactor cavity;
- The impossibility of locating the position of a considerable amount of fuel;
- Two detonations 1-2 sec. apart;
- An unnatural bright glow above the reactor cavity after the explosion;
- An isotope shift in the fuel samples studied, including the shift towards $^{235}U$ [10, 11];
- The attraction of electrical cables to steam pipes.

These facts do not fit into the accepted theories on the accident mechanism and have not been interpreted so far. The existence of such facts leads one to dig for the answer to the question of how did it come that this has happened.

Here, we have to recall major definitions in nuclear fission reactor physics. The intensity of neutron breeding in the core of reactor is described by the coefficient of neutron breeding **k** that is the ratio of neutron number of a certain generation to similar number in preceding generation. The excess reactivity $\rho$ is defined as **(k-1)/k**. For $\rho=0$ reactor is in a steady-state regime. For $\rho<0$ and $\rho>0$ the reactor, respectively, is slowing down and accelerating. The overwhelming part of neutrons produced by the decay of nuclei is born in fact instantaneously (namely, in a time much less than the lifetime of generation of neutrons, which is determined, in turn, by the capture and absorption of neutrons). A small fraction, $\beta$, is emitted by the products of the decay with a large enough delay, ~ 10 seconds, (these neutrons are called the delayed ones). For various types of reactors the $\beta$ value is varying in the range from 0.002 to 0.007. For the given reactor RBMK-1000, before the accident the $\beta$ value was 0.0045. The excess reactivity depends on parameters of the medium inside reactor, including the coolant density $\gamma$ ($\gamma$ is the ratio of the volume of the steam to the total volume of the coolant). The steam reactivity coefficient, $\alpha_\varphi$, is defined as a ratio of the rate of changing the reactivity $\rho$ to the rate of changing $\gamma$.

In the official version [12 - 16], an analysis of the origin and history of the accident was based on a numerical modeling. The authors suggested the following main sources of the accident:

- development of a strong inhomogeneity, in vertical direction, of energy release (i.e. of neutron density) in the reactor, because of downward motion of safety rods;
- positive value of the steam reactivity coefficient.

According to official version, the development of the accident proceeded in the following way: a localized source of power have caused a localized overheating of the coolant; this, in turn, resulted in a decrease of coolant density and, correspondingly, an increase of excess reactivity, The latter has resulted in the rise of neutron density and, respectively, power (note that the power of heat release is proportional to neutron density). Such a scenario may be considered as a sort of development of instability of the time behavior of neutron density.

We managed to show (Sec. 4) the time of development of such instability to substantially exceed the actual time of reactor's runaway during the accident, even if one takes, as an input data, the disputable curve «a» for the dependence of excess reactivity coefficient on the coolant density. Note that besides the steam reactivity coefficient, the temperature (i.e. Doppler effect-caused) and power coefficients may influence the dynamics of developing instability via diminishing the rate of reactivity rise up [18]. Just the latter makes it very difficult to attain the positive values of the excess reactivity, exceeding the $\beta$ value, in real experiments.

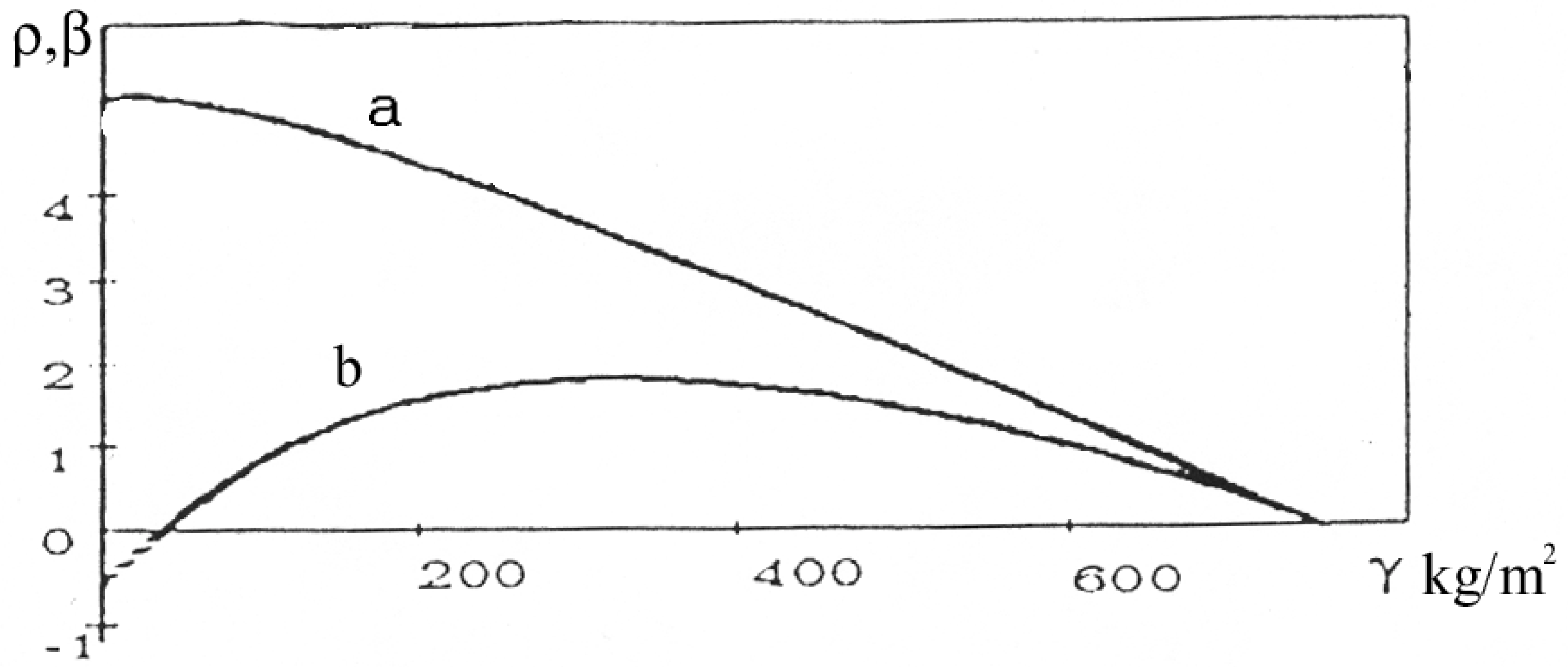

Fig. 1. The excess reactivity, in the units of $\beta$, as a function of the coolant density $\gamma$: «a» - the calculation [14] after the accident with the help of a specially modified numerical code; «b» - the projected curve [17] (i.e. calculated before the accident with the help of widely accepted numerical codes).

Regarding the cause of reactor's runaway, the authors [12 - 16] claim, using their specially modified numerical models, that the decrease of coolant density should be accompanied with a fast rise of reactor's excess reactivity up to the value of **$5\beta$**, see curve "a" in Fig. 1 taken from [14]. The curve "b" in Fig. 1 shows the dependence calculated at the stage of design [17], which agrees well with experimental results [18] obtained during the test of RBMK-1000 reactor.

It is seen that these curves substantially differ at low values of $\gamma$. The calculated curve «b» is confirmed by the results of experimental tests, whereas the curve «a» is based on a single event that took place in 1986 at the 4th unit of the Chernobyl Nuclear Power Plant (CNPP). No doubt the implications and importance of the accident of April 26, 1986 at CNPP are unprecedented; from scientific viewpoint, however, the curve «b» has a larger credit.

Note that for the values of the excess reactivity $\rho>\beta$, reactor acceleration is due to instantaneous neutrons. Therefore, the validity of the curve $\rho(\gamma)$ in [14] would imply that the reactor might be accelerated merely by complete withdrawal of the coolant. This, however, contradicts basic principles of reactor design in which the impossibility of reactor runaway under condition of complete withdrawal of the coolant is a key requirement.

## 3. THE DATA ON THE RUNAWAY OF THE REACTOR.

It has been shown rather convincingly [19] that the runaway of the high-power channel-type reactor (RBMK-1000) of the fourth generating unit of the Chernobyl Plant during the emergency involved the delayed neutrons. This conclusion [19] was based on instrument readings which indicated that the power acceleration during the first 6 sec. was developing at a constant value of the excess reactivity, **$\rho \sim 0.5\beta$,** and the time dependence of the power corresponded approximately to the law $N = 200e^{t/3}$ MW, with a period of 3 seconds. After 4 sec., a signal pointing to a sharp increase in the gas pressure in the reactor graphite stack was detected. Thus, in accordance with the available data, the acceleration of the reactor power lasted on the whole **$t \geq 10$** sec. Based on this fact, it was reasonably concluded [19] that the reactor runaway proceeded with participation of delayed neutrons. Indeed, in the case of instantaneous neutrons, the process would be approximately 100 times as fast and it would be absolutely impossible to follow it by means of instruments at the control panel.

The above conclusions [19] are supported indirectly by the absence of visual damages of the wall of the so-called casing of the reactor. This fact has been identified in 1990 by the staff of the «Complex Expedition» of the Kurchatov Institute: drilling of wells made it possible to survey of reactor's internal surface, with the help of periscope. The reactor cavity was found to be empty, i.e. the reactor itself has flown away only in vertical directions, i.e. up and down. Not only the visual deformations and the damages of internal surface of casing were absent but even the paint coating, which withstands, according to specifications, a temperature of **t ~ 500° C**, as been well preserved. The photograph of a fragment of the reactor cavity taken with a camcorder through the well drilled in the room # 605 through the biological shield is given in Fig. 2. Visual inspection revealed the signs of soot only in the southeast area of casing's internal surface. This makes the hypothesis for the fire in the reactor and the subsequent melting of the fuel to be very unlikely. The experts who investigated the fragments of the fuel arrived at the same conclusion [20].

## 4. ANALYSIS OF THE INFLUENCE OF STEAM REACTIVITY COEFFICIENT

According to official version, the rise of reactivity is caused by large values of steam coefficient. In this approach, because the rate of reactivity growth is proportional to neutron density, the neutron flux will grow substantially slower as compared with the case of a sudden change of reactivity considered in [19].

Assuming the validity of the dependence $\rho(\gamma)$ given by the curve «a» in Fig. 1, one may obtain the following limitation for the steam coefficient $\alpha_\varphi$:

$$\alpha_\varphi = \frac{d\rho}{d\gamma} < \frac{6\beta}{75\%} = 3.6 \cdot 10^{-4}\%^{-1},$$

because the data [18] (page 34) suggest that in the reactor which attains the steady-state regime of fuel reload, one has $\beta$**=0.0045**. Note that the official information [13] indicates even a smaller value $\alpha_\varphi = 2 \cdot 10^{-4}\,\%^{-1}$.

According to reactor design [18], the steam capacitance of the reactor at nominal power operation amounts **1.5 T/s**, with the average value of $\gamma$, at reactor's outlet, of 15% value and the amount of the coolant inside the reactor not less than **30 T**; the rate of variation in time of the coolant density (including that of steam) is proportional to the power (i.e. neutron density).

Eight Main Circulation Pumps (MCP) flow the coolant through the reactor. The schedule of the tests of the 4$^{th}$ power-generating unit of CNPP assumed four of these pumps to be fed by the electric circuit of the 3$^{rd}$ power-generating unit of CNPP. Therefore, these four pumps have been sufficient for reactor cooling, at least, up to 50% of nominal power whereas the reactor runaway has started from 6% of nominal power.

A relation would limit the reactivity growth, if the circulation of the coolant would be stopped completely (that is actually not possible though):

$$\frac{d\rho}{dt} = \frac{d\rho}{d\gamma} \cdot \frac{d\gamma}{dt} < \frac{6\beta}{0.75} \cdot \frac{1.5Ts^{-1}}{30T} \cdot \frac{W}{W_0},$$

where **W = 200 MW** is the initial value of reactor's power for the runaway process, $W_0$ **= 3200 MW** is the nominal power. This gives an equation for maximal rate of $\rho(t)$ growth:

$$\frac{d\rho}{dt} = \alpha \cdot \beta \cdot \frac{n}{n_0}, \qquad (1)$$

where $\alpha < 0.025\ s^{-1}$, n and $n_0$ are neutron density and its initial value. Note that Eq. (1) is valid locally as well because we assume the coolant that is being confined in a closed space to be evaporated due to released heat.

The equations of reactor's kinetics may be written in the form [21]:

$$\frac{dn}{dt} = \frac{\rho - \beta}{\mathrm{T}} n + \sum_i \lambda_i C_i$$
$$\frac{dC_i}{dt} = \frac{\beta_i n}{\mathrm{T}} - \lambda_i C_i, \qquad (2)$$

where $\mathbf{C_i}$ and $\boldsymbol{\lambda_i}$ are, respectively, the densities and inverse lifetimes of nuclei emitting the delayed neutrons, $\mathbf{T = 10^{-3}\ s}$, the lifetime of one generation of delayed neutrons. The solutions of above equations in the one-group approximation (with $\boldsymbol{\lambda}$**=0.1** $\mathbf{s^{-1}}$) for the cases of step-wise and linear time dependence of reactivity are given in [21]. The multi-group description makes the calculations cumbersome but doesn't change the conclusion qualitatively.

We analyze the system of nonlinear differential equations Eqs. (1) and (2) in one-group approximation for delayed neutrons. Introducing the following dimensionless variables:

$$\tilde{\rho} = \frac{\rho}{\beta}; \qquad \tilde{n} = \frac{n}{n_0};$$

one arrives at the following equations:

$$\frac{d\tilde{\rho}}{dt} = \alpha \cdot \tilde{n}$$
$$\left(\frac{\mathrm{T}}{\beta}\right)\frac{d^3\tilde{\rho}}{dt^3} + \left(\frac{\mathrm{T}\lambda}{\beta} + 1\right)\frac{d^2\tilde{\rho}}{dt^2} - \left(\frac{d\tilde{\rho}}{dt}\right)^2 - \tilde{\rho}\left(\frac{d^2\tilde{\rho}}{dt^2} + \lambda\frac{d\tilde{\rho}}{dt}\right) = 0 \qquad (3)$$

Integrating these equations, with allowing for initial conditions $\boldsymbol{\rho(0)=0}$, $\boldsymbol{\rho'_t(0)=\alpha}$ и $\boldsymbol{\rho''_{tt}(0)=\alpha \cdot n'_t(0)=0}$ (because runaway has started from the steady state), we obtain:

$$\left(\frac{\mathrm{T}}{\beta}\right)\frac{d^2\tilde{\rho}}{dt^2} + \left(\frac{\mathrm{T}\lambda}{\beta} + 1\right)\frac{d\tilde{\rho}}{dt} - \tilde{\rho}\frac{d\tilde{\rho}}{dt} - \frac{\lambda\tilde{\rho}^2}{2} = \left(\frac{\mathrm{T}\lambda}{\beta} + 1\right)\alpha \qquad (4)$$

As far as in our case $\mathbf{T \cdot \lambda \ll \beta}$ и $\mathbf{T \cdot \alpha \ll \beta}$, Eq. (4) takes the form:

$$e^{-\frac{\beta}{\mathrm{T}}t}\left(\frac{\mathrm{T}}{\beta}\right)\frac{d}{dt}\left[e^{\frac{\beta}{\mathrm{T}}t}\frac{d\tilde{\rho}}{dt}\right] = \alpha + \frac{1}{2}e^{-\lambda t}\frac{d}{dt}\left[e^{\lambda t}\tilde{\rho}^2\right] \qquad (5)$$

The nonlinear term in Eq. (5) contributes at values $\boldsymbol{\rho \sim (2 \cdot \alpha/\lambda)^{1/2} \sim 0.7}$. It is not difficult to solve these equations for $\mathbf{t \ll \alpha^{-1} = 40\ s}$. Coming back to variable of Eqs. (2), we obtain the following expansion up to cubic power terms:

$$\rho(t) = \beta \cdot \left[\alpha \cdot t + \frac{1}{2} \cdot \alpha^2 \cdot t^2 + \left(\frac{1}{2}\alpha^3 + \frac{\lambda}{6}\alpha^2\right) \cdot t^3 + o\left((\alpha t)^3\right)\right],$$
$$n(t) = n_0 \cdot \left[1 + \alpha \cdot t + \left(\frac{3}{2}\alpha^2 + \frac{\lambda}{2}\alpha\right) \cdot t^2 + o\left((\alpha t)^2\right)\right]. \qquad (6)$$

Recall that $\mathbf{\alpha < 0.025\ s^{-1}}$, and the solution of Eq. (6) is valid for time less that 40 seconds, i.e. at least for the first 10 seconds. This means that in the frame of adopted assumptions the rise of the power during the first 10 seconds cannot exceed noticeably a factor of 1.5. This agrees with the solutions [21] because in our case at first stage the excess reactivity increases linearly at the rate **α**. As far as during the accident the reactor increased its power by a factor of **exp** each 3 seconds, it follows that even the steam reactivity coefficient as high as taken in [12] (see the curve «a» in Fig. 1) can not lead to reactor runaway under condition of overheating of the coolant.

As mentioned, the multi-group model for describing the delayed neutrons doesn't change major results (the reader may check this as well): the increase of the power by a factor of **exp** from **200 MW** to **530 MW** would require not less than 20 seconds whereas actual rise up has happened in 3 seconds.

This rises a question about the validity of the calculated dependence of excess reactivity on coolant density (the curve «a» in Fig. 1) [14-16] which seem to be too high as compared with the respective projected values [17] calculated with the help of widely accepted numerical codes. We arrive at a conclusion that the proposed numerical models [14-16] seem to interpret the reactor runaway with recourse to disputable deformation of widely accepted numerical codes.

## 5. ANALYSIS OF SPATIAL INHOMOGENEITY

Let us consider the statement [12] about the role of spatial inhomogeneity of energy release (i.e. of neutron density) in reactor runaway. First, an analysis [22] of the post-accident survey of the reactor suggests that the bursting depressurization of reactor and flying reactor's away apart supports hypothesis for rather homogeneous increase of neutron density. Besides, spatial inhomogeneity under condition of a small positive excess reactivity was shown to suppress high spatial harmonics in the pioneering papers by Enrico Fermi [23]. For a large enough reactivity, the high spatial harmonics grow up, at least, lower that the fundamental one. Indeed, the first equation in Eq. (2) with allowing for spatial inhomogeneity takes the following form [23]:

$$\frac{dn}{dt} = D\Delta n + \frac{\rho - \beta}{\mathrm{T}} n + \sum_i \lambda_i C_i \qquad (7)$$

where $\mathbf{D = l \cdot v/3}$ is the coefficient of neutron diffusion, **v** - thermal velocity of the neutron, $\mathbf{l = (\sigma_s \cdot n)^{-1}}$ – mean free path with respect to scattering. Note also that $\mathbf{D = L^2 / T}$, where **T** as above the lifetime of one generation of instantaneous neutrons, $\mathbf{L = (l \cdot \Lambda)^{1/2}}$ - the diffusion length, $\mathbf{\Lambda = (\sigma_c \cdot n)^{-1}}$ - means free path with respect to capture and absorption. The diffusion length is **L ~ 50 cm** for graphite and ~ 3 cm for water. We seek for solutions of Eqs. (7) in the form **n(t) = φ(x,y,z)·f(t)** with the conditions **n = 0** at the reactor boundary. Thus we find:

$$\Delta\varphi = -k^2\varphi$$
$$\frac{dn}{dt} = \frac{\rho - (kL)^2 - \beta}{\mathrm{T}} n + \sum_i \lambda_i C_i \qquad (8)$$

It follows that in fact the inhomogeneity decreases the reactivity by the value $\mathbf{(2\pi L/a)^2}$, where **a** is the wavelength of spatial harmonic. In other words, the diffusion term that, due to spatial inhomogeneity of neutron distribution, appears in reactor kinetic equations «washes out» the fluctuations of neutron spatial distribution.

We can summarize the above considerations of Secs. 3, 4:

1. The official version of the accident contradicts the fundamentals of reactor physics in the following points:

- dependence of reactivity on coolant density is overestimated and disagrees with projected data calculated with the help of widely accepted numerical codes;
- assumption of a strong inhomogeneity of reactivity is not compatible with the phenomenology of the accident and the basic principles of a fission reactor laid by the pioneering papers by Fermi;
- our easily verifiable analytic estimates of the time period of instability's growth exceeds by far the respective results of a numerical modeling which was a basis for official conclusion;

2. As far as the convincing explanation of the accident origin is absent yet, the contemporary state of the science seems to be unable to provide such an explanation.

3. To interpret the origin of the accident, we suggest assuming existence of a new physical phenomenon (or even a number of such phenomena).

## 6. EXPERIMENTS ON STUDYING THE MAGNETIC MONOPOLE

Apart from the neutron mechanism, other mechanisms of uranium fission are also known to exist, for example, fission induced by slow muons [24]. . The mechanism of uranium fission under the action of magnetic monopoles has been considered theoretically [25]. It was suggested [25] that the monopole-nuclear interaction is so strong that a monopole that comes close to a nucleus can induce $^{238}$**U** fission.

To explain the experimentally observed facts, the formation of magnetic monopoles during a current interruption caused by electric discharge on a metallic foil in a fluid has been proposed as a working hypothesis [9]. In the authors' opinion, this hypothesis provides an explanation for the abnormal tracks recorded using nuclear emulsions, for the observed nuclear transformation, and for the shift of $^{57}$**Fe** Mossbauer spectra.

In order to confirm the hypothesized formation of monopoles, experiments on detection of the $^{238}$**U** fission induced by magnetic monopoles were carried out [26]. These experiments established that the original isotopic composition of uranium has changed towards $^{235}$**U** under the action of a "strange" radiation.

The decrease in the specific activity of uranium upon the electric discharge on a metallic foil noted in [26] is, most likely, related to the transformation of uranium nuclei. However, the fact that the monopoles predicted previously [6, 7] are leptons suggests that they should influence markedly the β-decay. Substantial distortions of the β-decay periods for the isomeric $^{234m}$**Th** nuclei, which are daughter products of $^{238}$**U**, were detected in experiments [26].

Thus, the experiments provided crucial arguments in favor of the existence of magnetic monopoles and substantial support for the hypotheses stated previously [9]. Let us assume that magnetic monopoles have entered the RBMK reactor and find out what would be the consequences, relying on the results of [9, 26].

## 7. THE IMPACT OF DISTURBED RATE OF DECAY OF NUCLEI-EMITTERS OF DELAYED NEUTRONS UPON REACTIVITY

In order to describe the impact of the disturbance of decay rates on reactor dynamics, one has to change a little the value of coefficients in reactor's kinetic equations. The nuclei-emitters of delayed neutrons in Eqs (2) are usually divided into the following six groups [21] (see Table 1). According to solutions of Eqs. (2), in the steady-state regime of reactor operation one has:

$$C_i = \beta_i / (\lambda_i T) n,$$

It follows that the number nuclei-emitters of delayed neutrons exceeds the number of instantaneous neutrons by two orders of magnitude.

| **i** | **$\lambda_i$, s$^{-1}$** | **$t_i$, s** | **$\beta_i$** |
|---|---|---|---|
| 1 | 14,0 | 0,071 | 0,00025 |
| 2 | 1,61 | 0,62 | 0,00084 |
| 3 | 0,456 | 2,19 | 0,0024 |
| 4 | 0,151 | 6,50 | 0,0021 |
| 5 | 0,0315 | 31,7 | 0,0017 |
| 6 | 0,0124 | 80,2 | 0,00026 |

Table 1. The parameters of nuclei emitting of delayed neutrons for the six-group representation of neutron kinetics.

The reactor is permanently rich with prolific amount of atoms whose nucleus is capable of emitting the neutrons. Therefore, the disturbance of the physical mechanism of nuclei decay with emission of delayed neutrons may result in a substantial change of neutron density. The matter is that all the sorts of nuclei emitting the delayed neutrons, which are forty in number [27, 28], decay via two channels, namely (i) β-decay and (ii) emission of a neutron, with the probability of the second channel rarely exceeding the value of 20%. The data on atoms emitting the delayed neutrons, taken from [27, 28], are presented in Table 2.

Equations (2) take into account the concentration of only those atom-emitters that were already decayed via the neutron channel whereas the atoms-emitters that underwent β-decay (i.e. β-decayed products) are considered to be lost for the chain reaction. In fact, the neutrons that contributed to the source of β-decayed products were taken into account in the increase of losses, i.e. in the decrease of reactivity ρ. Besides, a part of decays of atom-emitters is a source of similar atom-emitters. The allowance for all the above-mentioned issues gives us the following modification of Eqs. (2):

$$\frac{dn}{dt} = \frac{[\rho + (1-k_0)\beta_{tot}] - \beta_{tot}}{\mathrm{T}} n + \sum_i \lambda_i k_i C_i \,,$$

$$\frac{dC_i}{dt} = \frac{\beta_i}{\mathrm{T}} n - \lambda_i C_i \qquad (9)$$

where $\mathbf{k_i}$ is the probability of the neutron channel of atom-emitter's decay, $\boldsymbol{\beta_i}$ is a fraction of nuclei of the group **i** with allowance for those which are subject to β-decay, $\boldsymbol{\beta_{tot}}$ is a sum of $\boldsymbol{\beta_i}$**;** and $\mathbf{k_0}$ is the value of $\mathbf{k_i}$ averaged over distribution of atom-emitters before the accident. If the disturbance of decay process somewhat suppressed the β-decay, this may lead to a relative increase of the contribution of the neutron channel $\mathbf{k_i}$. As far as $\mathbf{k_i < 0.2}$ (see Table 2), there is rather large reserve for the increase of $\mathbf{k_i}$. Once the latter happened and the average value of $\mathbf{k_i}$ attained a value $\mathbf{k_1}$, the runaway time may be estimated, from Eq. (9), as follows:

$$\tau = \frac{2T}{\beta_0}\left[\sqrt{1 + 4\lambda\frac{T}{\beta_0}\cdot\frac{(k_1 - k_0)}{k_0}} - 1\right]^{-1}, \qquad (10)$$

where $\boldsymbol{\beta_0 = k_0 \cdot \beta_{tot}}$. For $\mathbf{T \cdot \lambda} \ll \boldsymbol{\beta}$ we have:

$$\tau = \lambda^{-1}\frac{k_0}{k_1 - k_0}.$$

It follows that for $\mathbf{k_1 \sim 4 \cdot k_0}$ the time of power increase by a factor of **exp** may be ~ 3 seconds. In other words, the increase of the role of neutron channel in the decay of atoms emitting the delayed neutrons results in an increase of the fraction of delayed neutrons in the core that, in turn, may result in the runaway of reactor.

## 8. A PHENOMENOLOGICAL MODEL OF NUCLEI TRANSMUTATION

The experimental results [9, 26] have shown the electric discharges on a metallic foil in a fluid to cause the transmutation of nuclei that distorts natural isotopic ratio for a number of chemical elements. The above transmutation was found to significantly differ from the known nuclear reactions: the input and output energies are small and do not exceed **10 keV** per newly born atom. The process of nuclear transmutation was suggested to be essentially a collective phenomenon, i.e. it cannot be reduced to a sum of a large number of pair collisions.

An attempt to interpret the observed phenomena have led to a numerical model, similar to that in [29], which is based on major conservation laws, namely those of:

- baryonic charge (i.e. the number of nucleons)
- electric charge;
- lepton number;

- energy,

Such an approach may find such a distribution of output nuclei that gives the minimal defect of total energy with respect to input distribution of nuclei. The binding energies of nuclei were taken from [30].

The model treats the elementary act of transmutation as such an interchange of nucleons within the cluster composed of input nuclei, which limits the defect of energy (per elementary act) to 10-30 keV. The calculation gives a distribution of output nuclei vs. that of input nuclei as resulted from an isotope transmutation process. This makes it possible to find the distributions whose energy differ from the input energy by a value as small as **~0.01 keV/atom**, i.e. these energies practically coincide because their difference coincides with the accuracy of measuring the binding energies of nuclei.

A comparison of experimental results with those of numerical modeling have shown [31] that the bigger is the size of the cluster in a model elementary act, the smaller is the gap between this theory and experiment. This supports the hypothesis for a collective nature of the process. Besides, it was found that for the input distributions composed of stable isotopes the closest output distributions are composed of stable isotopes as well. The latter agrees with experimental data.

It was suggested, on the basis of a numerical modeling, that the electric discharge on titanic foil doped with vanadium might produce the following nuclei:

- helium (the input sample does not contain helium);
- rare isotope $^{36}$**Ar** (natural abundance 0.34%);
- isotope $^{57}$**Fe** (natural abundance 2.2%).

Interestingly, just the production of $^{57}$**Fe** appeared to be the peculiar feature of vanadium as a dopant as compared with other chemical elements.

The experiments carried out with water solutions of vanadium salts confirmed [31] the above hypothesis in all three points that seem to be not occasional. Similar modeling of the transmutation for the case of uranium nuclei predicts the possibility of the shift of isotope ratio towards enrichments with $^{235}$**U**. Also, the modeling always predicts the production of large amount of hydrogen that is born essentially by the nuclear transmutation rather than hydrolysis. Therefore, the rise of reactivity could be caused not only by a change of the physical mechanism of nuclei emitting the delayed neutrons but also due to enrichment of the fuel with $^{235}$**U**.

## 9. DISTORTION OF ISOTOPE RATIO

The latter mechanism for the reactor runaway should result in a considerable isotopic shift of the $^{235}$**U** to $^{238}$**U** ratio due to fission of $^{238}$**U**. It is known that at the moment of the accident, the amount of $^{235}$**U** corresponded to an effective enrichment of 1.1 %. The uranium isotope ratios measured for soils in the area closely adjacent to the Power Plant, reported in [11], show an essential shift toward the enrichment in $^{235}$**U** (up to 27 %). These results were verified once again in the laboratories headed by Academician of the Ukrainian Academy of Sciences N. Shcherbak; the experiments confirmed the initial results [11].

Unfortunately, only the isolated groups have carried out the mass-spectrometry studies of the fuel from the 4-th unit of CNPP, during 10 years since the accident, and their results are not summarized. The results of experimental investigations of the dispersed phases of aerosols and secondary uranium minerals have been reported in [10]. It follows from the study that the observed $^{235}$**U** to $^{238}$**U** isotopic ratio corresponds to an enrichment of ~2 %; however it cannot be referred to a fresh fuel, as the $^{239}$**Pu**/$^{235}$**U** ratio is ~2.5, while, it should be 5 times lower. This discrepancy is too great to be regarded as being due to methodical errors. Although the results are scanty, they confirm the increase in isotopic ratio toward $^{235}$**U**, thus providing a strong argument supporting the stated hypothesis.

## 10. PROBABLE ROLE OF MAGNETIC MONOPOLE IN THE ACCIDENT

This brings about reasonable questions of what could be the source of magnetic monopoles at the $4^{th}$ unit of the Chernobyl power plant and how could they get into the reactor? The idea of invoking the magnetic charges as a mechanism of the Chernobyl accident has arisen during a study of the physical properties of the “strange" radiation observed in [26]. In experiments dealing with electric discharge on metallic foils in fluids [9] for nuclear emulsions and film detectors located at distances of up to **2 m** from the setup axis, abnormally broad tracks similar to a caterpillar's track appeared regularly. Since the sizes of the tracks did not allow one to explain their origin in terms of known kinds of radiation **($\alpha$, $\beta$, $\gamma$)**, the existence of a new type of radiation, conventionally referred to as "strange", was assumed. When a weak magnetic field, **$H_Z$ ~20 Oe**, was applied to the setup along the **Z**-axis, the pattern of the tracks changed. This circumstance suggested a magnetic nature for the detected radiation and provided grounds for regarding the radiation as a flux of magnetic particles [6, 7].

The current source used in the described experiments [9] was a discharge of a capacitor bank. In the tests done on April 26, 1986, the 8$^{th}$ turbine generator was disconnected from the substation and served as a power source for the purposes of only the 4$^{th}$ unit of the Chernobyl Power Plant. It is noteworthy that the initial power of the running under it's own momentum turbine generator was **40 MW** and the run lasted for ~40 sec.; hence, an occasional short in the electric circuit could have created conditions similar to those used in experiments in [26]. This analogy is largely intuitive but it well complies with evidence given by the operating personnel.

Yu. Tregub, the supervisor of the previous shift at the 4-th unit told the following. "First, I heard a characteristic noise of a shutting-down turbine generator. About six seconds later there was a stroke. I thought that the turbine blades were broken. Then another stroke followed. I looked at the upper floor and felt that it was going to fall down. I moved away to the safety shield. The instruments displayed a terrible emergency. I ran out of the building … a floodlight shone from the "Romashka" roof but some glow was also seen above the 4-th unit" [32].

R. Davletbayev, deputy supervisor of the turbine department said: "After several seconds, a low-pitched sound was heard from the turbine building, the floor and the walls were severely shaken, the dust and small-sized chips fell down from the ceiling, the fluorescent lighting died out, and there became darkish. A hollow stroke accompanied by thunder-like bursts was immediately heard. Then the lighting appeared again".

A.Dyatlov, the deputy chief engineer at the second tail of the Chernobyl Power Plant said: "I heard the first stroke from the turbine building. It was heavy but not as heavy as the next one, which was heard several seconds later. This was perceived as either one long stroke or two strokes following one another. The second one was more intensive" [33].

Thus, contrary to the generally accepted opinion, this hypothesis suggests that the development of the accident started at the turbine building and that pressing the emergency button AZ-5 incidentally coincided in time and by no means could prevent the disaster. The initial suggestion of the formation of magnetic monopoles at the moment of turbine generator run under it's own momentum can be advanced to form some scenario of the accident development. The magnetic monopoles, which have presumably formed in the vicinity of turbine generators, could get into steam pipes. Since oxygen is paramagnetic, magnetic particles should form so-called "bound states" with oxygen and move along the steam pipes, together with the steam, as in wave-guides. A "magnetic current" should have flown in the steam pipes. The electric wires located near such a field should be attracted to the magnetic current formed by the monopoles moving along the steam pipes. This can be really observed when one passes along the steam pipe route; moreover, some of the distribution boards were torn off together with the fastening fittings and fragments of partitions (near the separator department). In the separator buildings, even the partitions were ruined. The magnetic charges having got into the main circulation pumps should have caused a failure in the electric motor operations. Apparently, it is this fact that is responsible for the failure of power supply for four main circulation pumps (two north and two south ones). The failure took place exactly in those pumps that were supplied from the running under it's own momentum turbine generator # 8. The other four main circulation pumps were supplied from the third unit and these pumps remained intact.

After entering the reactor, the magnetic monopoles should have interacted both with the **$^{238}$U** nuclei and the nuclei emitting the delayed neutrons, which resulted in the growth of reactivity and hence, rise up of the power and the steam explosion. The probable production of a huge amount of hydrogen as resulted from nuclear transmutation [31] may cause the hydrogen explosion as well.

The two successive explosions in the region of the reactor at the moment of the accident [33] can be consistently explained within the framework of the mechanism we consider if one takes into account the difference between the pipeline lengths from the turbine building to the north and south separators.

Based on experimental results [9], it can be claimed that not only $^{238}$**U** nuclei but also some other nuclei, for example, $^{12}$**C**, could be transformed under certain conditions under the action of magnetic monopoles. Previously, this was predicted theoretically [4]. Thus, it can be suggested that once magnetic monopoles have got into the reactor, the reactor graphite should also undergo a transformation. In a study of the elemental composition of the post-accident fragments of graphite blocks from the 4$^{th}$ unit of the Power Plant, considerable islets of **Al, Si, Na**, and **U** were found within the graphite depth, although it is well known that highly pure graphite is used in reactors. This fact can serve as indirect evidence supporting the assumption about partial transformation of graphite.

A number of eyewitnesses including the members of the Government Commission have noted that the glow observed above the ruined reactor during the first days after the accident was unnaturally colored [32]. This fact can be easily explained within the framework of interaction of magnetic monopoles with excited atoms, which shifts the electronic levels of optical transitions [34, 35], giving rise to a color spectrum unusual for the human eye.

## 11. CONCLUSIONS

The author is aware of the fact that his hypothesis may provoke quite an explainable skepticism among the professionals. However, any hypothesis is admissible if it is able to explain some facts that do not fit in the framework of the existing views and predicts some other facts that can be verified experimentally.

The following studies are proposed for verifying the hypothesis in question:

1. A thorough determination of the isotopic composition of uranium in the fuel-containing masses (FCM);
2. Determination of the isotopic composition of the graphite units and carbon contained in the FCM (certainly, with allowance for the conducted campaign);
3. It is quite probable that radionuclides not characteristic of a uranium fuel cycle will be detected, because some $^{238}$**U** should have split under the action of the monopoles;
4. Fresh fuel assemblies were left in the central room and remained tight. If magnetic monopoles did actually participate in the accident, some of these could get into the fresh fuel and thus distort the initial isotope ratio toward $^{235}$**U**.
5. Finally, a direct experiment can be carried out, because magnetic monopoles should be stable particles such as electrons and one could attempt to detect them using nuclear emulsions. The tracks of magnetic charges are rather typical [9] and can be easily identified. The monopoles themselves can be "pulled out" by means of a current coil.

The authors acknowledge that the pioneering paper [19] by the Corresponding Member of Russian Academy of Science G. Kruzhilin was a starting point for their speculations about the origin of the reactor runaway.

The authors are grateful to the colleagues V.Vladimirov, A.Volkovich, V.Geras'ko, A.Kornejev, K.Checherov, and many others who worked self-deniably to collect, crumb by crumb, the reliable information concerning the post-accident state of the 4-th unit of Chernobyl Nuclear Power Plant.

The authors regard it as their duty to bow before the operating personnel of the 4$^{th}$ shift of the 4$^{th}$ unit of Chernobyl Nuclear Power Plant whose bravery and professionalism helped to avoid a larger disaster.

## REFERENCES

1. Dirac. P. A. M. Proc. Roy. Soc. 1931. Ser. A. V. 133. P. 60.
2. Polyakov A.M., Spectrum of particles in quantum field theory. Pis'ma ZhETF (JETP Lett.),. 1974. V. 20. No. 6. PP. 430-433.
3. Hooft G. Nucl. Phys., 1974. Ser. A. V. 133. P. 60.
4. Lipkin H. J.: Monoponucleosis – the wonderful things that monopoles can do to nuclei if they are there. Monopole'83. Proceedings of NATO advanced research workshop, Ann Arbor, MI, USA, 6-9 Oct. 1983, pp. 347-358.
5. Rubakov V.A. Superheavy magnetic monopoles and the decay of proton. Pis'ma ZhETF (JETP Lett.), 1981. V 33. #12. PP. 141-153.
6. Lochak G., in: Advanced Electromagnetism (Foundations, Theory and Applications), T.W. Barrett and D.M. Grimes ed., World Scientific Publishing Company, Singapore, 1995.
7. Lochak G.,Ann. Fond. L. de Broglie, 8 (1983) 345, 9 (1984) 5. International journal of Theoretical Physics, A. Blaquiere, S. Diner and G. Lochak ed., Springer, WIEN, N.Y., 1987.
8. Schwinger I., Phys. Rev., p. 144, 1087, (1966)
9. Urutskojev L.I., Liksonov V.I., Tsinoev V.G., Experimental detection of a "strange " radiation and transformations of chemical elements. Prikladnaya Fizika (Applied Physics, in Russian), 2000. V. 4. PP. 83-100.
10. Kuz'mina I.E., Lobach Yu.N., Nuclear fuel and peculiar features of aerosols in installation "Shelter". Atomic Energy. 1997. #1. PP. 39-44.
11. Sobotovich E.V., Chebanenko S.I., Isotope contents of uranium in soils of the near zone of Chernobyl Nuclear Power Plant. Physics Doklady, 1990, PP. 885-888.
12. The report by The government commission on studying the origin and circumstances of the accident at the Chernobyl Nuclear Power Plant. «The sources of and facts about the accident of April, 26, 1986, at the 4-th unit of the Chernobyl Nuclear Power Plant. The operations on handling the accident and mitigating its implications». pp. 12-32 (in Russian).
13. Information on the accident at the Chernobyl Nuclear Power Plant and its implications, submitted to IAEA. Atomic energy, 1986, vol. 61, # 5, pp. 302-320.
14. Adamov E.O., Vasinger V.V., Vasilevskii V.P., et. al., An estimate of qualitative effects of probable perturbations during the accident at CNPP. – In: The first international workshop on severe accidents and their implications. «Nauka», Moscow, 1990.
15. Adamov E.O., et. al., Analysis of first stage of the accident at 4-th unit of Chernobyl Nuclear Power Plant, Atomic Energy, 1988, vol. 64, # 1, pp. 24-28.
16. Afanasiev A.A., et. al., Analysis of the accident at Chernobyl Nuclear Power Plant with allowance for reactor core, Atomic Energy, 1994, vol. 77, # 2, pp. 87-93.
17. R&D Supplement to technical project RBMK, I.V. Kurchatov Atomic Energy Institute, internal ref. № 35-877, 1966.
18. Dollezhal N.A., Yemel'yanov I.Ya. Channel nuclear power reactor. «Atomizdat», Moscow, 1980, pp. 22-23, 34, 50, 96-97 (in Russian).
19. Kruzhilin G.I., On the features of explosion of the reactor «Reactor Big Power Channel 1000» at Chernobyl Nuclear Power Plant. Physics Doklady, 1997. V. 354. # 3. PP. 331-332.
20. Anderson E.B., Burakov B.E., Pazukhin Z.M., Did the fuel of the 4-th unit of Chernobyl Nuclear Power Plant melt? Radiochemistry, 1992, #5, pp. 155-158.
21. Shultz M.A., Control of nuclear reactors and power plants. Westinghouse Electric Corporation, Pittsburgh, 1955 (PP. 29-70 in the book translated to Russian by "Publishers of the Foreign Literature", Moscow, 1957).
22. Kiselev A.N., Surin A.I., Checherov K.P., Post-accident survey of reactor at 4-th unit of Chernobyl Nuclear Power Plant, Atomic Energy, 1996, vol. 80, # 4, pp. 240-247.
23. Fermi E., Research papers (in Russian), vol. 2, «Nauka», Moscow, 1972, pp. 316-326.
24. Belovitskii G.E., Rossel K., Instantaneous fission of nuclei of uranium by slow negative muons. Brief communications on physics (P.N. Lebedev Physics Institute), No. 9-10, 1996.

25. Fiorentini G. The coupling between magnetic charges and magnetic moments. Monopole'83. Proceedings of a NATO advanced research workshop, ANN arbor, MI, USA, 6-9 oct. 1983, pages 317-331.

26. Volkovich A.G., Govorun A.P., Gulyaev A.A., Zhukov S.V., Kuznetsov V.L., Rukhadze A.A., Steblevskii A.V., Urutskojev L.I., Observations of effects of isotope ratio distortions in Uranium and breakdown of secular distribution for Thorium-234 under condition of electric explosion. Brief communications on physics (P.N. Lebedev Physics Institute), 2002, #8, pp. 45-50.

27. Physical quantities. Handbook (Eds. Grigoriev I.S., Meilikhov E.Z.) «Energoatomizdat», Moscow, 1991 (in Russian).

28. Gangrskii Yu.P., Dalkhsuren B., Markov B.N., The products of nuclear splitting. «Energoatomizdat», Moscow, 1986.

29. Kuznetsov V.D., Myshynskii G.V., Zhemennik V.I., Arbuzov V.I., In: Materials of 8$^{th}$ Russian Conference on cold transmutation of nuclei of chemical elements. Moscow, 2001, pp. 308-332.

30. «The 1995 update to the atomic mass evaluation» by G. Audi and A.H. Wapstra, Nuclear Physics A595 vol. 4 p.409-480, December 25, 1995.

31. Filippov D.V., Urutskojev L.I., Gulyaev A.A., Klykov D.L., Dontsov Yu.P., Novosjolov B.V., Steblevskii A.V., Stolyarov V.L., A phenomenological model for low-energy transmutation of nuclei of chemical elements and its comparison with experiment (in press).

32. Chernobyl's reporting. «Planeta», Moscow, 1988 (in Russian).

33. Dyatlov A.S., «Chernobyl. As it was». «NauchTechIzdat», Moscow, 2000 (in Russian).

34. Sidney D. Drell, Norman M. Kroll, Mark T. Mueller, Stephen J. Parke, Malvin A. Ruderman «Energy loss of slowly moving magnetic monopoles in matter», Physical Review Letters, v.50, number 9, p 644-649.

35. D. Lynden-Bell, M. Nouri-Zonoz "Classical monopoles: Newton, NTU space, gravitation lens and atom specters", Review Modern Physics, Vol. 70, No. 2, April 1998, p. 421-445.

| Z | isotope | $T_{1/2}$, s. | k, % |
|---|---|---|---|
| 37 | Rb-98 | 0.12 | 15.0 |
| 37 | Rb-97 | 0.17 | 30.0 |
| 37 | Rb-96 | 0.20 | 14.2 |
| 36 | Kr-94 | 0.21 | 2.2 |
| 55 | Cs-147 | 0.21 | 25.0 |
| 34 | Se-91 | 0.27 | 21.0 |
| 55 | Cs-146 | 0.34 | 13.4 |
| 35 | Br-92 | 0.36 | 22.0 |
| 37 | Rb-95 | 0.38 | 8.9 |
| 34 | Se-89 | 0.41 | 5.0 |
| 53 | I-141 | 0.47 | 39.0 |
| 35 | Br-91 | 0.54 | 10.8 |
| 55 | Cs-145 | 0.59 | 14.3 |
| 38 | Sr-99 | 0.60 | 3.4 |
| 53 | I-140 | 0.60 | 22.0 |
| 51 | Sb-136 | 0.82 | 23.0 |
| 33 | As-86 | 0.90 | 10.5 |
| 55 | Cs-144 | 1.00 | 3.0 |
| 50 | Sn-134 | 1.04 | 17.0 |
| 39 | Y-97 | 1.13 | 1.6 |
| 54 | Xe-142 | 1.24 | 0.4 |
| 36 | Kr-93 | 1.29 | 2.1 |
| 39 | Y-99 | 1.40 | 1.2 |
| 52 | Te-138 | 1.40 | 5.6 |

| Z | isotope | $T_{1/2}$, s. | k, % |
|---|---|---|---|
| 34 | Se-88 | 1.52 | 0.2 |
| 51 | Sb-135 | 1.71 | 13.9 |
| 55 | Cs-142 | 1.71 | 0.2 |
| 54 | Xe-141 | 1.73 | 0.0 |
| 55 | Cs-143 | 1.78 | 1.8 |
| 36 | Kr-92 | 1.85 | 0.0 |
| 35 | Br-90 | 1.92 | 21.2 |
| 33 | As-85 | 2.03 | 23.0 |
| 53 | I-139 | 2.38 | 10.2 |
| 37 | Rb-94 | 2.76 | 10.6 |
| 52 | Te-137 | 2.80 | 2.2 |
| 35 | Br-89 | 4.38 | 13.5 |
| 37 | Rb-92 | 4.50 | 0.0 |
| 33 | As-84 | 5.60 | 0.1 |
| 34 | Se-87 | 5.60 | 0.2 |
| 37 | Rb-93 | 5.85 | 1.4 |
| 53 | I-138 | 6.53 | 2.6 |
| 51 | Sb-134 | 10.40 | 0.1 |
| 35 | Br-88 | 16.00 | 6.9 |
| 52 | Te-136 | 17.50 | 0.9 |
| 53 | I-137 | 24.50 | 6.7 |
| 55 | Cs-141 | 24.90 | 0.1 |
| 35 | Br-87 | 55.60 | 2.4 |

Table 2. The emitters of delayed neutrons (k is the fraction of the decay with emission of the neutron, $T_{1/2}$ is the half-life).

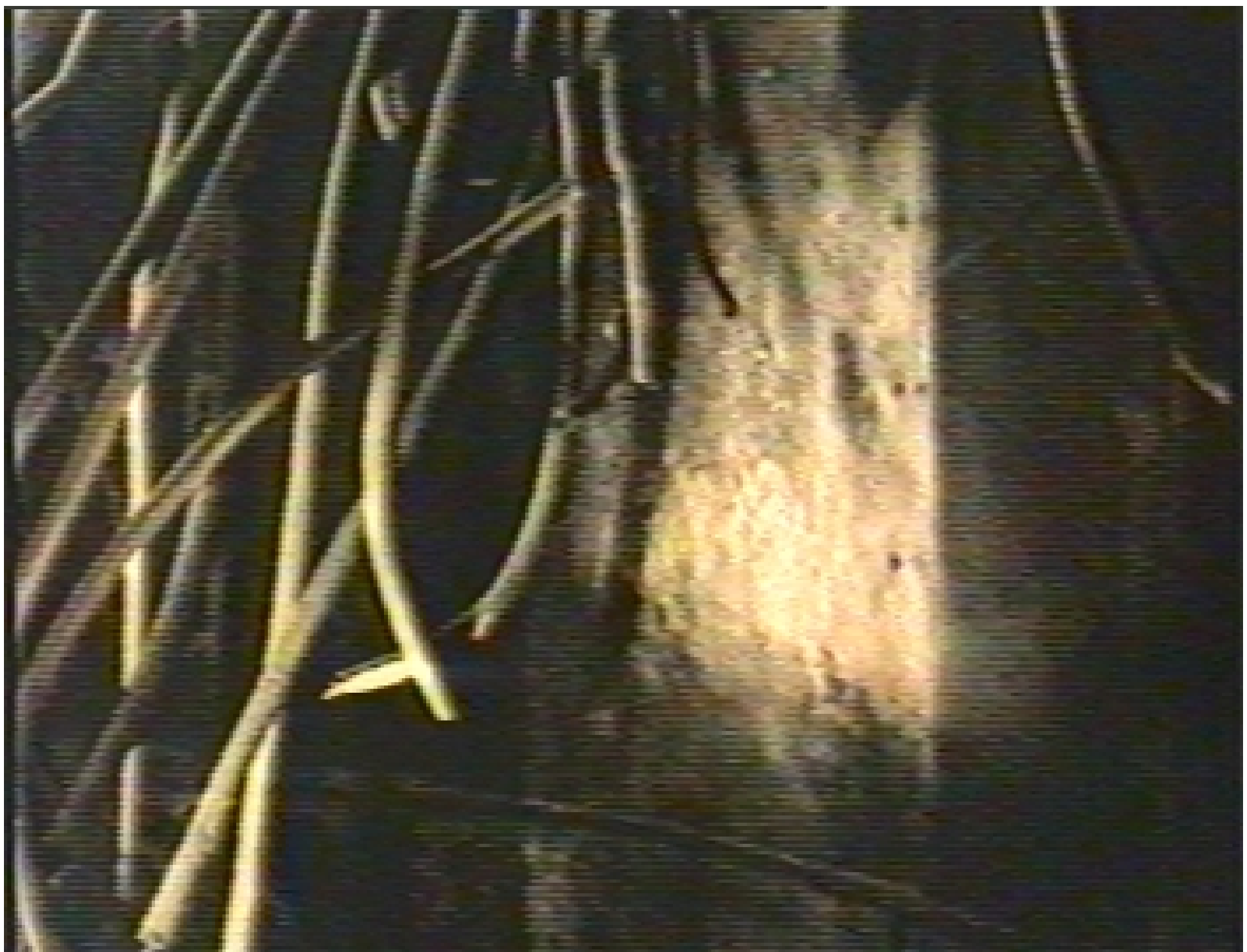

Fig. 2. A fragment of the internal surface of the reactor shaft.